\documentclass{article}
\usepackage{amssymb}
\usepackage{epsfig}
\usepackage{amsmath}

\input{tcilatex}

\begin{document}

\title{\textbf{Compact calculation of the Perihelion Precession of Mercury
in General Relativity, the  Cosmological Constant and Jacobi's Inversion problem.}}
\author{G. V. Kraniotis \footnote{georgios.kraniotis@physik.hu-berlin.de} \\
Humboldt Universit$\rm{\ddot a}$t zu Berlin,\\
Mathematisch-Naturwissenschaftliche Fakult$\rm{\ddot a}$t I\\
Institute f$\rm {\ddot u}$r Physik, \\
Newtonstra$\ss$e 15, D-12489 Berlin, \\
Germany \footnote{
HU-EP-03/20, May 2003}, \\
 \and S. B. Whitehouse \footnote{steve.whitehouse@CSIconsulting.co.uk} \\
Royal Holloway, University of London , \\
Physics Department, \\
Egham Surrey TW20-0EX, U.K.}
\maketitle

\begin{abstract}
The geodesic equations resulting from the Schwarzschild gravitational metric 
element are solved exactly including the contribution from the Cosmological 
constant. The exact solution is given by genus 2 Siegelsche modular forms.
For zero cosmological constant the hyperelliptic curve degenerates into an 
elliptic curve and the resulting geodesic is solved by the Weierstra$\ss$ 
Jacobi modular form. The solution is applied to the precise calculation of 
the perihelion precession of the orbit of planet Mercury around the Sun.
\end{abstract}

\bigskip

\bigskip \newpage

\section{ Introduction}

\subsection{\protect\bigskip Motivation}

Mercury is the inner most of the four terrestrial planets in the 
Solar system, moving with high velocity in the Sun's gravitational 
field. Only comets and asteroids approach the Sun closer at 
perihelion. This is why Mercury  offers unique possibilities for testing 
General Relativity \cite{Albert} and exploring the limits of alternative theories of 
gravitation with an interesting accuracy \cite{Pireaux}.

As seen from the Earth the precession of Mercury's orbit is measured to be 
5599.7 seconds of arc per century \cite{Will}. 
As observed, in 1859 by Urbain Jean-Joseph Leverrier there is a 
deviation of Mercury's orbit 
from Newtonian's predictions that could not be due to the presence of 
other planets. Urbain Jean-Joseph Leverrier had also been the 
first to attempt 
to explain this effect.

Perihelion precessions of Mercury and other bodies have been the subject 
of experimental study from AD 1765 up to the present. In 1882  Simon 
Newcomb  obtained the value 43 seconds 
per century for the discrepancy for Mercury \cite{Newcomb}.
According to Pireaux et al \cite{Pireaux}, the observed advance 
of the perihelion of 
Mercury that is unexplained by Newtonian planetary perturbations or
solar oblatness is
\footnote{ The observations seem to exclude  
Brans-Dicke theory with $\omega \sim 5$ whose 
post-Newtonian contribution to the perihelion shift would thus have been 
$39$ arc-seconds per century \ref{Mercobs}) \cite{Will, Pireaux}.}
\begin{eqnarray}
\Delta_{\omega_{obs}}&=&42.980\pm 0.002\; {\rm arc-seconds\; per\; century}
\nonumber \\
&=&\frac{2 \pi (3.31636\pm 0.00015)\times 10^{-5} {\rm{radians}}}{
415.2019 {\rm revolutions}} \nonumber \\
&=& 2\pi (7.98734\pm 0.00037)\times 10^{-8} {\rm radians/revolution}
\label{Mercobs}
\end{eqnarray}

It is the purpose of this paper to provide the first $precise$ 
calculation of the orbit and the perihelion 
precession of Mercury in General Relativity including cosmological constant 
contributions \cite{ALVERTOS, GVKSBW}.  We also compare the result
with 
observations (\ref{Mercobs}). We assume that the motion of Mercury is a time-like geodesic 
in a Schwarzschild space-time surrounding the Sun.

Bounds on the magnitude  of the cosmological constant from 
the observed value of Mercury's perihelion shift have been estimated using 
different methods of approximation by Islam \cite{ISLAM} and 
Cardona et al  \cite{Cardona} with conflicting results.  
The upper bounds obtained were respectively, $|\Lambda| \leq 10^{-42}{\rm cm^{-2}}$
\cite{ISLAM} and 
$|\Lambda| \leq 10^{-55}{\rm cm^{-2}}$ \cite{Cardona}. 
The latter value 
compares favourably to the values for the magnitute of the cosmological 
constant obtained from large-scale observations \cite{PERLMUTTER} and cosmological considerations \cite{GVKSBW}.

In \cite{GVKSBW} a direct connection between general exact solutions of 
general relativity with a cosmological constant in large-scale cosmology and 
the theory of modular forms and elliptic curves \cite{wiles,DON} 
was established.

As we shall see in the main body of the paper, 
the two-body central orbit problem under consideration is 
very interesting mathematically
and becomes more involved with the inclusion of the cosmological constant.
In particular, the integration of the 
geodesic orbital equations involves the {\em inversion problem}  
of Abelian hyperelliptic integrals
and the resulting solution which was first studied by outstanding 
mathematicians  such as Jacobi, Abel, 
Weierstra$\ss$, Riemann, G$\rm{\ddot o}$pel, Rosenhain and Baker.
The exact treatment of the orbital problem  and the techniques
developed in this paper are general and should be of interest for a
variety of problems in various 
fields of cosmology which  are discussed in the main text.

The material of this paper is organized as follows. In the rest of the introduction 
we review planned satellite experiments that are relevant
to the study of Mercury and the minor planets. In section 2, 
starting with the  Schwarzschild metric,  and including the cosmological
constant, we derive the time-like  geodesic equations.
In section 3, we solve exactly the geodesic differential equations and
calculate the precession  of
the perihelion of Mercury as well as the perihelion and aphelion 
distances of the planet, with and without the cosmological constant.
Our approach involved the solution of Jacobi's inversion problem i) for an 
elliptic integral (case without cosmological constant)  and ii) for a genus-2  
hyperelliptic integral (in the general case with $\Lambda \not =0$) . 
Here we acknowledge the contribution by Whittaker of the exact
solution of time-like geodesics in the Schwarzschild field with vanishing
cosmological constant \cite{WHITAKKER}. However, in his account
no attempt was made in calculating the
physical characteristics of the orbit of planet Mercury.
For an alternative method of computation of perihelion precession in
the case of vanishing cosmological constant we refer the reader to ref.\cite{KERNER}.
In order to make the 
presentation self-contained we include in the same 
section 
a comprehensive presentation of the inversion problem.
In section four we present a discussion and a summary of our main
results as well as further possible applications. Finally, in an
appendix we present the definitions of the genus-2 theta functions
that solve Jacobi's Inversion problem.

\subsection{Planned Satellite Experiments}
Solar and planetary astronomers and cosmologists are still fascinated with
the precession of the Perihelion of Mercury and the dynamics and properties
of near earth objects such as the minor planets (i.e. Icarus). In fact, there
is a growing interest in this exciting and vibrant field with several
satellite missions planned for the next few years. Listed below are some of
the key satellite experiments (see Pireaux et. al. \cite{Pireaux} for an
overview):

\begin{itemize}
\item GAIA (ESA) \ \ \ \ \ \ \ \qquad

planned launch 2009, key goal to provide observational data to tackle an
enormous range of important problems related to to the origin, structure and
evolutionary history of our Galaxy. The data will create a precise three -
dimensional map of more than one billion stars throughout our Galaxy - \ a
stellar census. GAIA will observe and discover several hundred thousand
minor planets, determine their physical properties and orbital
characteristics. Also three earth crossing asteroids, Icarus, Talos and
Phaeton, with favourable combination of distance and eccentricity, will
provide data to more accurately determine the PPN parameters giving upper
limits on the relativistic and solar quadrupole contributions to the
perihelion precessions of solar objects enabling a more stringent test of
general relativity. The GAIA mission is a hugely ambitious and exciting project
with a multitude of objectives (see\ GAIA Study Report\ \cite{GAIA}\ for
details).

\item BepiColombo (ESA) \ \ \ 

planned launch 2010, Mercury Orbiter Experiment, key objective to study the
Mercury environment in order to understand the formation and evolution of of
planets within the solar system, especially the four closest to the Sun -
Mercury, Venus, Earth and Mars. BepiColombo will measure Mercury's motion
more accurately than ever before, providing data that will allow the
determination of the precession of the perihelion, thus providing one of the
most rigorous tests of general relativity \cite{BepiColombo}.

\item Messenger (NASA)

planned launch 2004, a Mercury Orbiter experiment whose main goal is to
study the chacteristics and environment of Mercury in order to better
understand the fundamental forces that have formed it and the other
terrestrial planets \cite{Messenger}

\item Picard (C.N.E.S.) \qquad\ \ \ 

planned launch end of 2005, main object is to measure the Sun's quadrupole
moment contribution to the precession of the Perihelion of Mercury, thereby
offering another exacting test of general relativity \cite{Picard}
\end{itemize}

\qquad

Also there are several near Earth Satellites presently operational, LAGEOS,
LAGEOS II, CHAMP and GRACE. The combined data from these satellites will
enable the accurate measurement of the perigee advance in the Earth's
gravitational field (the determination of the PPN parameters $\gamma $, $%
\beta $, in the Earth's gravitational field), providing yet another test of
general relativity \cite{Iorio}.

All of these satellite experiments are expected, to various
degrees, to provide data that will provide stringent tests for general
relativity and to more accurately determine solar and planetary geophysical
characteristics.

Our precise results that follow should be of interest to all of the
above experiments.

\section{Schwarzschild solution and time-like geodesics}
Einstein's equations with the cosmological constant $\Lambda$ are as follows
\begin{equation}
R_{\mu\nu}-\frac{1}{2}g_{\mu\nu} R=(8\pi G/c^4)T_{\mu\nu}+\Lambda g_{\mu\nu}
\end{equation}
where $R_{\mu\nu},R$ denotes the Ricci tensor and scalar respectively.
Also $G$ denotes Newton's gravitational constant and $c$ the velocity of 
light. For vanishing stress-energy momentum tensor $T_{\mu\nu}$ the 
field equations reduced to 
\begin{equation}
R_{\mu\nu}=\Lambda g_{\mu\nu}
\end{equation}

The motion of a planet according to General Relativity is a time-like 
geodesic in a Schwarzschild space-time \cite{Karl} surrounding the Sun. 
The Schwarzschild solution for the metric taking into account the 
cosmological constant (see e.g., ref.\cite{OHANIAN,ISLAM}) is 
\begin{eqnarray}
ds^2&=&c^2 (1-\frac{2GM_{\bigodot}}{c^2 r}+\frac{1}{3}\Lambda r^2)dt^2 \nonumber \\
&-&(1-\frac{2GM_{\bigodot}}{c^2 r}+\frac{1}{3}\Lambda r^2)^{-1} dr^2 \nonumber \\
&-& r^2 (d\theta^2+\sin^2 \theta d\phi^2)
\label{schwarz}
\end{eqnarray}
where $M_{\bigodot}$ denotes the mass of the Sun.
For zero cosmological constant Eq.(\ref{schwarz}) reduces to the original 
 Schwarzschild metric element \cite{Karl}.
We note at this point, that by taking the time-like geodesic equation
as a starting point for the subsequent calculus represents an
approximation to the real situation:
it is strictly true for  point-like bodies with negligible
mass. However, since $m_M \ll M_{\bigodot}$ where $m_M$ denotes the mass
of Mercury the  approximation is adequate for all practical purposes. 

The geodesic equation has the general form:
\begin{equation}
\frac{d^2 x^{\mu}}{ds^2}+\Gamma^{\mu}_{\alpha\beta}\frac{dx^{\alpha}}{ds}
\frac{dx^{\beta}}{ds}=0
\label{timegeod}
\end{equation}

It is easily shown that the motion is confined to the equatorial plane 
$\theta=\frac{\pi}{2}$, so that (\ref{timegeod}) with $\mu=2$ 
is trivially satisfied 
\cite{OHANIAN}.
Furthermore, we can ignore (\ref{timegeod}) with $\mu=1$ in favour of 
(\ref{schwarz}), which is a first integral of the geodesic equations. 
The resulting equations can be written as
\begin{eqnarray}
1&=&g_{\mu\nu}\frac{dx^{\mu}}{ds}\frac{dx^{\nu}}{ds} \nonumber \\
&=&c^2 \left(1-\frac{2 G M_{\bigodot}}{c^2 r}+\frac{1}{3}\Lambda r^2\right)\left(
\frac{dt}{ds}\right)^2 \nonumber \\
&-&\left( 1-\frac{2 G M_{\bigodot}}{c^2 r}+\frac{1}{3}\Lambda r^2\right)^{-1}
\left(\frac{dr}{ds}\right)^2-r^2\left(\frac{d\phi}{ds}\right)^2 
\label{ar1}\\
&&\frac{d^2\phi}{ds^2}+\frac{2}{r}\frac{dr}{ds}\frac{d\phi}{ds}=0 
\label{ar2}\\
&&\frac{d^2 t}{ds^2}+\frac{2 (\frac{ G M_{\bigodot}}{c^2 }+\frac{1}{3}\Lambda r^3)}{r(r-2\frac{ G M_{\bigodot}}{c^2 }+\frac{1}{3}\Lambda r^3)}\frac{dr}{ds}\frac{dt}{ds}=0
\label{ar3}
\end{eqnarray}

Next, define $v:=dt/ds$ and $w:=d\phi/ds$. Then eqs(\ref{ar2}),(\ref{ar3}) 
can be written as
\begin{eqnarray}
\frac{dw}{dr}&+&\frac{2}{r}w=0 \\
\frac{dv}{dr}&+&\left[-\frac{1}{r}+\frac{1+\Lambda r^2}{r-2\frac{ G M_{\bigodot}}{c^2 }+
\frac{1}{3}\Lambda r^3}\right]v=0
\end{eqnarray}
which can be integrated to yield \cite{OHANIAN}
\begin{eqnarray}
w&=&d\phi/ds=\frac{L}{r^2} \\
v&=&\frac{dt}{ds}=\frac{{\cal E}}{1-\frac{2 G M_{\bigodot}}{c^2 r}+\frac{1}{3}\Lambda r^2}
\end{eqnarray}
where $L,{\cal E}$ are arbitrary constants.
Substituting in (\ref{ar1}) we get the following equation after definining 
a new variable $u:=r^{-1}$
\begin{equation}
\left(\frac{du}{d\phi}\right)^2=\frac{2 G M_{\bigodot}}{c^2} u^3-
u^2+\frac{2 G M_{\bigodot}}{c^2 L^2} u-\frac{1}{3}\left(\frac{1}{u^2
L^2}+1\right)\Lambda+\left(\frac{E^{\prime 2}-1}{L^2}\right)
\label{Geodesic}
\end{equation}

\section{Exact solution of the time-like geodesic and the perihelion 
precession}
\subsection{Precise calculation of the perihelion advance assuming zero 
Cosmological Constant.}
In the previous section we derived the geodesic equation for the orbital 
motion of the planet Mercury around the Schwarzschild gravitational
field of the
Sun given by eq.(\ref{Geodesic}).
In this section we will solve Eq.(\ref{Geodesic}) with a zero 			Cosmological  Constant, the non-zero, more general case,
will be considered in section 3.3.
Making the following definitions,
\begin{equation}
\alpha_{S}:=\frac{2 G M_{\bigodot}}{c^2},\;\beta:=
\frac{2 G M_{\bigodot}}{c^2 L^2}, \gamma:=-\frac{1-E^{\prime 2}}{L^2}
\label{defini}
\end{equation}
our geodesic equation reduces to:
\begin{equation}
\Biggl(\frac{du}{d \phi}\Biggr)^2=\alpha_{S}u^3-u^2+\frac{\alpha_{S}}
{L^2} u+\gamma
\label{geo2}
\end{equation}
Differentiating Eq.(\ref{geo2}) with respect to $\phi$ and comparing with the 
Newtonian term we get $cL=L_M$, where $L_M$ is the angular momentum 
per unit mass of the planet \cite{ISLAM}.
Equation (\ref{geo2}) is a cubic equation which can be reduced to the 
Weierstra$\ss$ form using the substitution:
\begin{equation}
u=\frac{4}{\alpha_S}U+\frac{1}{3\alpha_S}
\label{change}
\end{equation}
Then the Weierstra$\ss$ representation of the geodesic equation
becomes \footnote{The differential equation that the Weierstra$\ss$ 
function satisfies is the equation of an elliptic curve and is 
given by : $(\wp^{\prime}(z))^2=4\wp(z)^3-g_2 \wp(z)-g_3$, 
$\wp^{\prime}(z)=\frac{\partial \wp(z)}{\partial z}$, see \cite{GVKSBW}
for further details. The inversion of the elliptic integral 
 $ \int_{}^U \frac{dU}{\sqrt{4 U^3-g_2 U-g_3}}=\phi$, 
by the Weierstra$\ss$ function, i.e. $U=\wp(\phi+\epsilon)$,
is a simple case of the problem of inversion for elliptic
integrals.}
\begin{equation}
\Biggl(\frac{dU}{d \phi}\Biggr)^2=4 \; U^3 +\{-\frac{1}{12}+\frac{1}{4 L^2}
\alpha_S^2\} \; U-\frac{1}{216}+\frac{\alpha_S^2}{48 L^2}+\frac{\alpha_S^2}{16}
\gamma
\label{KW}
\end{equation}
where the Weierstra$\ss$ cubic invariants are given by:
\begin{eqnarray}
g_2&=&\frac{1}{12}-\frac{(2 G M_{\bigodot}/c^2)^2}{4 L_M^2} c^2
\nonumber \\
    &=&\frac{1}{12}-\frac{1}{4}{\cal L} c^2 \nonumber \\
g_3&=&\frac{1}{216}-\frac{(2 G M_{\bigodot}/c^2)^2}{48 L_M^2}c^2-
\frac{(2 G M_{\bigodot}/c^2)^2}{16}\left(\frac{E^2-c^2}{L_M^2}\right)
\nonumber \\
&=& \frac{1}{216}-\frac{1}{48}{\cal L} c^2-\frac{1}{16}{\cal L} (E^2-c^2)
\end{eqnarray}
where ${\cal L}:=\frac{\alpha_S^2}{L^2_M}$ and $E=cE^{\prime}$.
Then the solution of Eq.(\ref{KW}) is given by:
\begin{equation}
U=\wp(\phi+\epsilon)
\label{inveellip}
\end{equation}
and $\epsilon$ is a constant of integration.
The solution in terms of the original variables is given by:
\begin{equation}
u=\frac{1}{r}=\frac{4}{\alpha_S}\wp(\phi+\epsilon)+\frac{1}{3 \alpha_S}
\end{equation}
or 
\begin{equation}
r=\frac{\alpha_S}{4 \wp(\phi+\epsilon) +\frac{1}{3}}
\label{orbitnol}
\end{equation}

For the calculation of the perihelion precesion and the orbital 
characteristics of Mercury we use the following
values for the physical constants:
\begin{equation}
c=299\; 792 \;458\; {\rm m\; s^{-1}}\;\; , \; \alpha_S=2.953\; 250\; 08\; {\rm Km}
\label{lightvacspeed}
\end{equation}
The data is taken from \cite{Petley} and \cite{WasGB}, respectively.
The value of the speed of light in vacuum, given in
Eq.(\ref{lightvacspeed}), 
is exact, since the meter is the length of the path travelled by light
in vacuum during a time interval of $1/299\;792 \;458\;$ of a second \cite{Petley}.
As free parameters we may use ${\cal L}$ and $E$. Then
$L_M^2=c^2L^2=\frac{1}{{\cal L}} \alpha_S^2$  \footnote{The integration constant $L$ has dimensions of
length and thus scales with $\alpha_S$.}. Our free parameters  are mixed through the 
Weierstra$\ss$ invariants $g_2,g_3$ with exact numbers.

We will show in the next
section that there is a small parameter space for ${\cal L}$ and $E$ that reproduces
the observed values for the characteristics of
Mercury's orbit. Non-physical solutions that predict complex, negative or
too small orbital radii and precession values are disregarded.
The sign of the discriminant $\Delta=g_2^3-27 g_3^2$ determines the 
roots $e_i,i=1,2,3$ of the elliptic curve: $\Delta>0$, corresponds
to three real roots $e_i$ while for $\Delta<0$ two roots are complex 
conjugates and the third is real. In the degenerate case $\Delta=0$, 
(where at least two roots coincide) the elliptic curve becomes singular and 
the solution is not given by 
modular functions.

We find that the physically acceptable solutions that reproduce the 
orbital data of Mercury correspond to the 
case where $\Delta >0$, $e_3\rightarrow -\frac{1}{12}, e_2 \rightarrow -
\frac{1}{12}$  and $e_1\rightarrow 1/6$.
The analytic expressions for the roots of the elliptic curve may be
obtained by using the algorithm of Tartaglia and Cardano \cite{TARTACARDA}.
Their expressions are given in appendix B.
The Weierstra$\ss$ function, $\wp(z)$, is an even meromorphic elliptic 
function of periods $ 2\omega, 2\omega^{\prime}$ (i.e.,$\;\wp(z+2\omega)=
\wp(z)=\wp(z+2\omega^{\prime}), $ for all complex numbers $z$).
The two half-periods $\omega$ and $\omega^{\prime}$ are given by the 
following Abelian integrals (for $\Delta>0$) \cite{WHITAKKER}:
\begin{equation}
\omega=\int_{e_1}^{\infty}\frac{dt}{\sqrt{4t^3-g_2t-g_3}}, 
\;\;\;\omega^{\prime}=i\int_{-\infty}^{e_3}\frac{dt}{\sqrt{-4t^3+g_2t+g_3}}
\end{equation}
The value of the Weierstra$\ss$ function at the half-periods are the three roots of the cubic.
For positive discriminant $\Delta$ one half-period is real while the 
second is imaginary.
The exact expression for the precesion of the perihelion  
of planet Mercury, is given by 
\begin{equation}
\Delta_{\omega}^{GTR}=2(\omega-\pi)
\label{Precession}
\end{equation}
which is proportional to the deviation of the real half-period 
$\omega$ of the Weierstra$\ss$ 
modular form, from the transcendental number $\pi$.

The exact expressions for the minimum distance of planet Mercury from the Sun
($Perihelion$) and its maximum distance ($Aphelion$) are given by
\footnote{We organize the roots as:$\;\;e_1>e_2>e_3$}:
\begin{eqnarray}
r_P\equiv r_{Perihelion}&=&\frac{\alpha_s}{4 e_2+\frac{1}{3}} \\
r_A\equiv r_{Aphelion}&=& \frac{\alpha_s}{4 e_3+\frac{1}{3}}
\label{Perihelion}
\end{eqnarray}

For such orbits the radius vector oscillates between $r_A$ and $r_P$,
as the argument of the Weierstra$\ss$ function travels along the
straight line from $\omega^{\prime}$ to $\omega^{\prime}+\omega$ and
then to $\omega^{\prime}+2 \omega$ in
the fundamental period region \cite{Silverman}.

We also note the following. Although by construction the roots of the 
cubic are calculated with arbitrary precision, our output for the
perihelion and aphelion distances can only be displayed with nine
significant figures given the nine digit accuracy of the Schwarzschild
length $\alpha_S$. However, their ratio which is given by:
\begin{equation}
\frac{r_P}{r_A}=\frac{4 e_3+1/3}{4 e_2+1/3}
\label{goldenratio}
\end{equation}
constitutes a genuine and precise prediction \footnote{In
Newtonian theory the orbit of a planet is described by an ellipse with
eccentricity $e$ and semi-major axis $a$. For an ellipse, the
perihelion ($r_P^N$) and aphelion ($r_A^N$) distances, are $a(1-e), a(1+e)$, respectively,
and then $e=\frac{1-r_P^N/r_A^N}{1+r_P^N/r_A^N}$.}.

For a given choice of values for the free parameters,
equations (\ref{Precession})-(\ref{goldenratio}) are
the output of the {\em precise theory} assuming zero cosmological 
constant, for the corresponding physical 
quantities,  that should be tested against 
observations.

\subsection{Theoretical results}

The exact solution of the geodesic equations of the General Theory of
Relativity outlined in the previous section predicts a set of
coupled results,

\begin{tabular}{ll}
$\Delta \omega ^{GTR}$ & - Perihelion Precession \\
$r_{P}$ & - distance of closest approach \\
$r_{A}$ & - distance of farthest approach \\
\end{tabular}

based solely upon the values chosen for the ${\cal L}$ and $E$.
As we mentioned not all of input values for the free parameters  
provide us with physical solutions, or produce realistic data for the
orbit of Mercury. In appendix B, we provide some examples of solutions 
that correspond to specific choices of the free parameters that do not
describe the orbit of Mercury. More specifically, we describe
solutions for the motion of the test particle that are described by
special elliptic curves for which one of the Weierstra$\ss$
invariants, $g_2,g_3$,
vanishes \footnote{These elliptic curves have the property 
of complex multiplication, see \cite{GVKSBW} for the 
definition.}. In addition, we describe choices  of the initial
conditions that lead to a vanishing discriminant $\Delta$ (singular
elliptic curves).

For ${\cal L}=1.18554647\times 10^{-28}{\rm cm^{-2} \;s^2}$ $(L^2_M=7.35668\times 10^{38} \;{\rm cm^4}\;{s^{-2}})$ and 
$E=0.0299792454176\times 10^{12}\;{\rm cm\; s^{-1}}$
we obtain 
\begin{eqnarray}
e_2&=&-0.0833333172749111339,\;e_3=-0.083333322753852543388 \nonumber \\
\Delta_{\omega}^{GTR}&=&43.0017{\rm \frac{arcsec}{century}} \nonumber \\
r_{P}&=&4.59766539\times 10^{12}{\rm cm},\nonumber \\
r_{A}&=&6.97872168\times 10^{12} {\rm cm}
\end{eqnarray}
while the two  periods are: $\omega=3.141592904646,\omega^{\prime}=20.40864976\; i$
and the period ratio $\tau=6.496 i$, 
and the radius vector oscillates between $\frac{\alpha_s}{4 e_2+\frac{1}{3}}
\leq r \leq \frac{\alpha_s}{4 e_3+\frac{1}{3}}$.
For ${\cal L}= 1.1848820116975453\times 10^{-28}\;{\rm
cm^{-2}}\;{s^{2}}  $,
($L^2_M=7.36080550\times 10^{38}\;{\rm cm^4}\;{s^{-2}}$) and 
$E=0.0299792454178\times 10^{12}\;{\rm cm \;  s^{-1}}$ 
we obtain, $\Delta_{\omega}^{GTR}=42.9776{\rm
\frac{arcsec}{century}}$ with the half-periods, 
$\omega=3.141592904505435,\omega^{\prime}=20.409634\; i$
and the period ratio $\tau=6.497 i$. Also $r_P=4.60057668\times
10^{12} $cm, $r_A=6.98186816\times 10^{12}$cm.

In Tables \ref{EINSTEIN1}-\ref{EINSTEIN3} we list the predicted 
results for $\Delta_{\omega}^{GTR}, r_P,r_A$ keeping $E$ fixed at the
indicated value and varying ${\cal L}$. 
We see that by increasing the input value 
for ${\cal L}$ one can obtain a slightly larger value for the precession 
of perihelion. 
The parameter $E$ fixes 
the sum of the radii $r_P+r_A$. In fact, in tables
\ref{EINSTEIN1}-\ref{EINSTEIN3} the sum $\frac{r_P+r_A}{2}$ is equal
to $5.7909175\times 10^{12}$cm.
The latter value is reported as the data for the semi-major axis $a$
in ref.\cite{COX}.
In table \ref{EINSTEIN2} the value for $L_M$ that corresponds to the 
choice for ${\cal L}$ for fixed $\alpha_S$ is equal to the  
Newtonian value that would be obtained by assuming the orbit is an
ellipse with eccentricity $e$ \cite{OHANIAN,ISLAM}.

The choice of initial conditions presented in Tables
\ref{EINSTEIN1}-\ref{EINSTEIN3}, lead to a set of predictions for the 
perihelion advance in agreement with  observations in
Eq.(\ref{Mercobs}). In addition, they lead to a 
set of predictions for the perihelion and aphelion
distances of the planet, whose half-sum, $\frac{r_P+r_A}{2}$, is in
agreement with the data for the semi-major axis $a$
in ref.\cite{COX}. The calculated theoretical values for the
perihelion and aphelion distances 
compare favourably 
with best current values for the orbital data for Mercury \cite{NASA}. 
More precise measurements of the orbit of Mercury as described for instance in the ESA mission BeriColombo, will determine 
the perihelion and aphelion distances as well as the precession of
the perihelion more accurately and  will further
restrict the choice of initial conditions.

\begin{table}
\begin{center}
\begin{tabular}{|c|c|c|}\hline\hline
{\bf parameters} & {\bf roots} & {\bf predicted results} \\
${\cal L}=$
& $0.16666664004116$ &  $\Delta_{\omega}^{GTR}=42.9817{\rm
\frac{arc-sec}{century}}$ \\
$ 1.1849947026647969\times 10^{-28}\;{\rm cm^{-2}}\;{s^{2}}$ 
& $-0.083333317282230892$ & 
$r_{P}=4.59976206\times 10^{12}{\rm cm}$ \\
$E=$& $-0.083333322758930472$ &  $r_{A}=6.98207293\times 10^{12} {\rm cm}$ 
\\
$0.029979245417779875\times 10^{12}\;{\rm cm \; s^{-1}}$ & .& . \\
\hline \hline
\end{tabular}
\end{center}
\caption{Predictions for $\Delta_{\omega}^{GTR},r_P,r_A$ 
for the indicated choice for ${\cal L},E$. The two  half-periods are: $\omega=3.14159290452929,\omega^{\prime}=20.409059\; i$
and the period ratio $\tau=6.496 i$. Also $L^2_M=7.36010550\times 10^{38}\;{\rm cm^4}\;{s^{-2}}$}
\label{EINSTEIN1}
\end{table}

\begin{table}
\begin{center}
\begin{tabular}{|c|c|c|}\hline\hline
{\bf parameters} & {\bf roots} & {\bf predicted results} \\
${\cal L}=$ & $0.16666664004188$ &  $\Delta_{\omega}^{GTR}=42.9805{\rm
\frac{arc-sec}{century}}$ \\
$1.1849627128268641\times 10^{-28}\;{\rm
cm^{-2}}\;{s^{2}} $ 
& $-0.08333331728350096$ & 
$r_{P}=4.60012605\times 10^{12}{\rm cm}$ \\
$E=$& $-0.08333332275837917$ &  $r_{A}=6.98170894\times 10^{12} {\rm cm}$ 
\\
$0.029979245417779875\times 10^{12}\;{\rm cm\; s^{-1}}$ & .& . \\
\hline \hline
\end{tabular}
\end{center}
\caption{Predictions for $\Delta_{\omega}^{GTR},r_P,r_A$ 
for the indicated choice for ${\cal L},E$. The two  half-periods are: $\omega=3.141592904522524,\omega^{\prime}=20.409391\; i$
and the period ratio $\tau=6.4965 i$. Also $L^2_M=7.36030420\times 10^{38}\;{\rm
cm^4}\;{s^{-2}} $}
\label{EINSTEIN2}
\end{table}

\begin{table}
\begin{center}
\begin{tabular}{|c|c|c|}\hline\hline
{\bf parameters} & {\bf roots} & {\bf predicted results} \\
${\cal L}=$ & $0.166666640043693405$ &  $\Delta_{\omega}^{GTR}=42.9776{\rm
\frac{arc-sec}{century}}$ \\
$ 1.1848820116975453\times 10^{-28}\;{\rm cm^{-2}}\;{s^{2}}  $ 
& $-0.083333317286706146$ & 
$r_{P}=4.60104489\times 10^{12}{\rm cm}$ \\
$E=$& $-0.083333322756987258$ &  $r_{A}=6.98079010\times 10^{12} {\rm cm}$ 
\\
$0.029979245417779875\times 10^{12}\;{\rm cm\; s^{-1}}$ & .& . \\
\hline \hline
\end{tabular}
\end{center}
\caption{Predictions for $\Delta_{\omega}^{GTR},r_P,r_A$ 
for the indicated choice for ${\cal L},E$. The two  half-periods are: $\omega=3.141592904505435,\omega^{\prime}=20.410231661\; i$
and the period ratio $\tau=6.497 i$. Also 
$L^2_M=7.36080550\times 10^{38}\;{\rm cm^4}\;{s^{-2}}$. }
\label{EINSTEIN3}
\end{table}

\subsection {Precise calculation of the perihelion advance with the
contribution of the
Cosmological Constant.}

As we saw in section 3.1, the integration of the geodesic equation 
that describes the orbit of Mercury in the central gravitational 
field of the Sun (assuming vanishing cosmological constant) involved
an elliptic integral, whose inversion by the Weierstra$\ss$ elliptic 
function, provided the exact solution 
in closed analytic form for the orbit of the planet eq.(\ref{inveellip}).

Elliptic integrals are special cases of the so called Abelian integrals.
According to Jacobi an Abelian integral is an integral of the form $\int 
R(x,y)dx$, where $R(x,y)$ is a rational function in $x$ and $y$ and $x,y$ are 
connected through an equation $f(x,y)=0$, where $f$ is an irreducible polynomial. In the special case $y^2=P(x)$, where $P(x)$ is a polynomial of $n^{th}$ degree, with no multiple roots, the Abelian integral is called elliptic when $n=3$ or $n=4$, and hyperelliptic when $n\geq 5$.

Including cosmological constant effects we need to calculate the integral:
\begin{equation}
\int_{}^u \frac{du}{\sqrt{\frac{2 G M_{\bigodot}}{c^2} u^3-
u^2+\frac{2 G M_{\bigodot}}{c^2 L^2} u-\frac{1-E^{\prime 2}}{L^2}-
\frac{\Lambda}{3 L^2 u^2}-\frac{\Lambda}{3}}}=\phi
\label{hyperelliptic}
\end{equation} 
or 
\begin{equation}
\int_{}^u \frac{u\;du}{\sqrt{\frac{2 G M_{\bigodot}}{c^2} u^5-
u^4+\frac{2 G M_{\bigodot}}{c^2 L^2} u^3-\frac{1-E^{\prime 2}}{L^2} u^2-
\frac{\Lambda}{3 L^2 }-\frac{\Lambda}{3}u^2}}=\phi
\label{hyperelliptic1}
\end{equation}

Eq.(\ref{hyperelliptic}) defines a {\it hyperelliptic integral} whose 
inversion involves genus 2 Abelian-Siegelsche modular functions. 
The problem of inversion (whose solution was culminated in the 
Jacobisches Umkehrtheorem) of hyperelliptic integrals were first 
investigated by Abel \cite{Abel}, Jacobi, G$\rm{\ddot o}$pel and Rosenhain \cite{
Jacobi}, \cite{Adolph}, \cite{Georg}. The explicit solution of Jacobi's inversion problem 
in terms of higher genus 
theta functions was provided by G$\rm{\ddot o}$pel and Rosenhain for the 
case $n=5$ or 6, and the  general solution for the hyperelliptic case (i.e. 
 $\forall\; n \;\geq 5$) was provided by Weierstra$\ss$ \cite{KWeierstrass}. 
Riemann introduced the idea of a Riemann surface to study algebraic
singularities. He also introduced the Riemann theta function which
served 
as a useful tool for solving the Jacobi's inversion problem \cite{BerGeorgR}.
For full details and extended bibliography to the 
original literature we refer the reader to the book 
of Baker \cite{BAKER}. 

In what follows we discuss first the Jacobi's inversion problem and
then we proceed to determine the effect of the cosmological constant 
on the perihelion advance of Mercury as well as its effect on the
radii  $r_P,r_A$.

\subsection{Abel's theorem and Jacobi's inversion problem}

Let the genus $g$  Riemann hyperelliptic surface be described by the 
equation:
\begin{equation}
y^2=4 (x-a_1)\cdots (x-a_g) (x-c) (x-c_1)\cdots (x-c_g)
\label{Riemann}
\end{equation}
For $g=2$ the above hyperelliptic Riemann algebraic equation reduces to:
\begin{equation}
y^2=4 (x-a_1) (x-a_2) (x-c) (x-c_1) (x-c_2)
\label{Riemann}
\end{equation}
where $a_1,a_2,c_,c_1,c_2$ denote the finite branch points of the surface.

The Jacobi's inversion problem involves  finding the  solutions,
for $x_i$ in terms of $u_i$,  
for the following system of equations of Abelian integrals \cite{BAKER}:
\begin{eqnarray}
u_1^{x_1,a_1} & +&\cdots + u_1^{x_g,a_g} \equiv  u_1 \nonumber \\
\vdots        &+ &\cdots +  \vdots \;\;\;\;\;\;\;\;\;\;\;\;\;  \vdots \nonumber \\
u_g^{x_1,a_1}  &+&\cdots + u_g^{x_g,a_g} \equiv u_g
\end{eqnarray}
where 
$u_1^{x,\mu}=\int_{\mu}^{x}\frac{dx}{y},
u_2^{x,\mu}=\int_{\mu}^{x}\frac{x dx}{y},
\cdots,u_g^{x,\mu}=\int_{\mu}^{x}\frac{x^{g-1} dx}{y}$.

For $g=2$ the above system of equations takes the form:
\begin{eqnarray}
\int_{a_1}^{x_1} \frac{dx}{y}+\int_{a_2}^{x_2} \frac{dx}{y}\equiv u_1 \nonumber \\
\int_{a_1}^{x_1} \frac{x\;dx}{y}+\int_{a_2}^{x_2} \frac{x\;dx}{y} \equiv u_2
\label{Umkehr}
\end{eqnarray}
where $u_1,u_2$ are arbitrary.
The solution \footnote{The definitions of the genus two theta functions appearing in the solution 
of Jacobi's inversion problem are given in the appendix A.
} is given by the five equations \cite{BAKER} 
\begin{eqnarray}
\frac{ \theta^2(u|u^{b,a})}{
\theta^2(u)}&=&A(b-x_1)(b-x_2)\cdots(b-x_g) \nonumber \\
&=&A(b-x_1)(b-x_2) \nonumber \\
&=&\pm \frac{(b-x_1)(b-x_2)}{\sqrt{e^{\pi i P P^{\prime}}f^{\prime}(b)}};
\label{inveb}
\end{eqnarray}
where $f(x)=(x-a_1)(x-a_2)(x-c)(x-c_1)(x-c_2)$,
and $e^{\pi i P P^{\prime}}=\pm 1$ according as $u^{b,a}$ is an odd or even 
half-period. Also $b$ denotes a finite branch point and the branch
place $a$ being at infinity \cite{BAKER}.
The symbol $\theta(u|u^{b,a})$ denotes a genus 2 theta function with
characteristics: $\theta(u;q,q^{\prime})$ \cite{BAKER}, where $u,=(u_1,u_2)$, 
denotes two independent variables, see appendix A for further details.
From any $2$ of these equations, eq.(\ref{inveb}), the upper 
integration bounds $x_1,x_2$ 
of the system of differential equations eq.(\ref{Umkehr})
can be expressed as single valued 
functions of the arbitrary arguments $u_1,u_2$.
For instance,
\begin{equation}
x_1=a_1+\frac{1}{A_1 (x_2-a_1) } \frac{\theta^2(u|u^{a_1,a})}{\theta^2(u)}
\label{inve1}
\end{equation}
and 
\begin{eqnarray}
x_2&=&-\;\frac{\Bigl[(a_2-a_1)(a_2+a_1)+\frac{1}{A_1}\frac{\theta^2(u|u^{a_1,a})}{\theta^2(u)}-\frac{1}{A_2}\frac{\theta^2(u|u^{a_2,a})}{\theta^2(u)}\Bigr]}
{2 (a_1-a_2)} \nonumber \\
&\pm&\frac{\sqrt{\Bigl[(a_2-a_1)(a_2+a_1)+
\frac{1}{A_1}\frac{\theta^2(u|u^{a_1,a})}{\theta^2(u)}-\frac{1}{A_2}\frac{\theta^2(u|u^{a_2,a})}{\theta^2(u)}\Bigr]^2-
4(a_1-a_2)\eta}}{2(a_1-a_2)} \nonumber \\
\label{inve2}
\end{eqnarray}
where 
\begin{equation}
\eta:=a_2\;a_1(a_1-a_2)-\frac{a_2}{A_1}\frac{\theta^2(u|u^{a_1,a})}{\theta^2(u)}+\frac{a_1}{A_2}\frac{\theta^2(u|u^{a_2,a})}{\theta^2(u)}
\end{equation}
Also, $A_i=\pm \frac{1}{\sqrt{e^{\pi i P P^{\prime}}f^{\prime}(a_i)}}$.

The solution can be reexpressed in terms of generalized Weierstra$\ss$ functions:
\begin{equation}
x_k^{(1,2)}=\frac{\wp_{2,2}(u)\pm \sqrt{\wp^2_{2,2}(u)+4\wp_{2,1}(u)}}{2},
\;\;k=1,2
\end{equation}
where
\begin{equation}
\wp_{2,2}(u)=\frac{(a_1-a_2)(a_2+a_1)-\frac{1}{A_1}\frac{\theta^2(u|u^{a_1,a})}{\theta^2(u)}+\frac{1}{A_2}\frac{\theta^2(u|u^{a_2,a})}{\theta^2(u)}}{a_1-a_2}
\end{equation}
and 
\begin{equation}
\wp_{2,1}(u)=\frac{-a_1a_2(a_1-a_2)-\frac{a_1}{A_2}\frac{\theta^2(u|u^{a_2,a})}{\theta^2(u)}+\frac{a_2}{A_1}\frac{\theta^2(u|u^{a_1,a})}{\theta^2(u)}}{a_1-a_2}
\end{equation}
Thus, $x_1,x_2$, that solve Jacobi's inversion problem Eq.(\ref{Umkehr}), 
are solutions of a quadratic equation \cite{Jacobi,BAKER}
\begin{equation}
U x^2-U^{\prime} x+U^{\prime\prime}=0
\end{equation}
where $U,U^{\prime},U^{\prime\prime}$ are functions of $u_1,u_2$.
In the particular case that the coefficient of $x^5$ in the quintic 
polynomial is equal to 4, $U=1,U^{\prime}=\wp_{2,2}(u),U^{\prime\prime}=
\wp_{2,1}(u)$.

\subsection{Inversion of the hyperelliptic integral.}
Let us apply the solution of the Jacobi's inversion problem given
by eqs.(\ref{inveb})-(\ref{inve2}) to the hyperelliptic integral 
(\ref{hyperelliptic1}).
We have the following correspondence between the variables 
of  the hyperelliptic integral and the variables in (\ref{Umkehr}):
the upper integration bounds in (\ref{Umkehr}) $x_i$, correspond to 
$u_i$ and the periods $u_i$ correspond to  $\phi_i$.
Then for:
\begin{eqnarray}
\phi_1&=&\phi_1^{u,a_2} \nonumber \\
\phi_2&=&\phi_2^{u,a_2}
\end{eqnarray}
\begin{equation}
u_1=a_1,\;\;u_2=u,\; u\;\; {\rm arbitrary}.
\end{equation}
Thus we have that:
\begin{eqnarray}
\frac{1}{A_1}\frac{\theta^2\left(\Phi|\Phi^{a_1,a}\right)}{\theta^2(\Phi)}
&=&(u_1-a_1)(u_2-a_1)=0 \nonumber \\
\frac{1}{A_2}\frac{\theta^2\left(\Phi|\Phi^{a_2,a}\right)}{\theta^2(\Phi)}
&=&(a_1-a_2)(u-a_2)
\end{eqnarray}
and the solution for $u$ in the hyperelliptic integral $\phi_2=\phi_2^{u,a_2}=\int_{a_2}^{u} \frac{udu}
{y}$ (where $y$ is the radical of the quintic polynomial \footnote{
we can always bring the quintic polynomial of the hyperelliptic 
integral into the defining form (\ref{Riemann}) of the genus-2 
hyperelliptic surface.}) is given by
\begin{equation}
u_2=u=\frac{\wp_{2,2}(\Phi)}{2}+\frac{\sqrt{\wp_{2,2}(\Phi)^2+4\wp_{2,1}(\Phi)}}{2}
\end{equation}
where 
\begin{equation}
\wp_{2,2}(\Phi)=a_1+a_2+\frac{1}{A_2 (a_1-a_2)}\frac{\theta^2(
\Phi|\Phi^{a_2,a})}{\theta^2(\Phi)}
\end{equation}
and 
\begin{equation} 
\wp_{2,1}(\Phi)=-a_1
a_2-\frac{a_1}{A_2(a_1-a_2)}\frac{\theta^2(\Phi|\Phi^{a_2,a})}{\theta^2(\Phi)}
\end{equation}

\subsection{Roots of the quintic, periods of the hyperelliptic surface
and the cosmological constant effect}

In principle, the sign of the cosmological constant is an additional
parameter, besides its magnitude.
Recent observations of large-scale cosmology indicate a positive 
cosmological constant of magnitude $\sim 10^{-56}{\rm cm^{-2}}$ 
\cite{PERLMUTTER}.

In the presence of the cosmological term there are five branch points 
for the hyperelliptic surface, eq.(\ref{Riemann}), which are obtained by solving
the quintic polynomial that appears in the time-like geodesics, Eq.( \ref{hyperelliptic}).

For negative cosmological constant all the roots are real. 
For positive cosmological constant and depending on its magnitude 
and the values of the parameters ${\cal L},E$ three roots are real and two complex conjugates. For some particular values 
all the roots are real.

Let us start our discussion with the case of the negative sign.
Let us arrange the roots in $ascending$ order of magnitude and denote them
by $e_{2g},e_{2g-1},\cdots,{e_0},\;g=2$, so that $e_{2i},e_{2i-1}$, 
are respectively $c_{g-i+1},a_{g-i+1}$ and $e_0$ is $c$. We then define
linearly  independent Abelian integrals of the first kind
\cite{BAKER}, 
denoted by $U_i^{x,a},\;i=1,\cdots g$, whose periods we want to calculate. These integrals are
such that $dU_r^{x,a}/dx=\psi_r/y$, where $\psi_r$ is an integral 
polynomial in $x$, of degree $g-1=1$ at most, with only real
coefficients.
Then the half-periods, 
$U_r^{e_4,e_3}$ and $U_r^{e_2,e_1}$, are $real$, while the half-periods
$U_r^{e_3,e_2}$ and $U_r^{e_1,e_0}$ are purely $imaginary$ \cite{BAKER}. 
For clarity, by $U_2^{e_4,e_3}$ we denote $\int_{e_3}^{e_4}\frac{
x dx}{y}$, $U_1^{e_4,e_3}$ denotes $\int_{e_3}^{e_4}\frac{
 dx}{y}$ and similarly for the rest of the periods.

Every other period can be expressed as a linear combination of the above 
Abelian integrals.
For instance, the period $U_r^{e_1,a}$ is given by the equation:
\begin{equation}
U_r^{e_1,a}=-U_r^{e_2,e_1}-U_r^{e_4,e_3}+U_r^{e_1,e_0}
\end{equation}
while the period $U_r^{e_0,a}=U_r^{e_0,\infty}$ is given by
\begin{equation}
U_r^{e_0,a}=-U_r^{e_2,e_1}-U_r^{e_4,e_3}
\end{equation}

For $\Lambda=-10^{-55}{\rm cm^{-2}}, 
{\cal L}= 1.1848820116975453\times 10^{-28}\;{\rm
cm^{-2}}\;{s^{2}}  (L^2_M=7.3608055\times 10^{38}\;{\rm cm^4}\;{s^{-2}}),
E=0.0299792454178\times 10^{12}{\rm cm\; s^{-1}}$ the roots are:
\begin{eqnarray}
e_4&=&-1.143376827586711663025\times 10^{-24}  \nonumber \\
e_3&=&1.14337682758671162302505\times 10^{-24}  \nonumber \\
e_2&=&1.43228141354887622443346\times 10^{-13}   \nonumber \\
e_1&=&2.173640544970253310548427\times 10^{-13}  \nonumber \\
e_0&=&0.000003386099606939
\end{eqnarray}

The half-periods are calculated to be:
\begin{eqnarray}
U_2^{e_3,e_2}&=&2.264048243121482\;i,\; U_2^{e_1,e_0}=20.4094269\;i \nonumber \\
U_2^{e_1,e_2}&=&3.14159290450544
\end{eqnarray}
the real half period: $\int_{e_2}^{e_1}\frac{udu}{y}=3.14159290450544$, 
leads to $\Delta_{\omega}^{GTR}=42.9776 {\rm \frac{arc-sec}{century}}$ in agreement 
with observations (\ref{Mercobs}).

For  $\Lambda=-10^{-42}{\rm cm^{-2}}, 
{\cal L}= 1.1848820116975453\times 10^{-28}\;{\rm
cm^{-2}}\;{s^{2}},
E=0.0299792454178\times 10^{12}{\rm cm\; s^{-1}}$ the roots are:

\begin{eqnarray}
e_4&=&-3.615599293152843540392\times 10^{-18}  \nonumber \\
e_3&=&3.6157507116827139496388\times 10^{-18}  \nonumber \\
e_2&=&1.4322814108727319240989\times 10^{-13}   \nonumber \\
e_1&=&2.1736405461322123121789\times 10^{-13}  \nonumber \\
e_0&=&0.000003386099606939
\end{eqnarray}

The periods are calculated to be:
\begin{eqnarray}
U_2^{e_3,e_2}&=&2.26404824460\;i,\; U_2^{e_1,e_0}=20.4094269\;i \nonumber \\
U_2^{e_1,e_2}&=&3.14159290255,\;U_2^{e_3,e_4} \in R
\end{eqnarray}
Here there is substantial effect on the perihelion advance due to the 
cosmological constant with  $\Delta_{\omega}^{GTR}=42.6427{\rm \frac{arc-sec}{century}}$ in conflict 
with observations (\ref{Mercobs}).

For $\Lambda=10^{-42}{\rm cm^{-2}}, {\cal L}= 1.1848820116975453\times 10^{-28}\;{\rm
cm^{-2}}\;{s^{2}},
E=0.0299792454178\times 10^{12}{\rm cm\; s^{-1}}$ the roots are:
\begin{eqnarray}
e_1&=&0.000000010574713017622\times \frac{4}{\alpha_s} \nonumber \\
e_2&=&0.0000000160482602747\times \frac{4}{\alpha_s} \nonumber \\
e_3&=&0.2499999733770\times \frac{4}{\alpha_s} \nonumber \\
e_4&=&(-5.58970979303625\times
10^{-18}-2.6694981178048\times10^{-13}i)\times \frac{4}{\alpha_s}
\nonumber \\
e_5&=&\bar{e}_4
\end{eqnarray}
with the periods $\int_{e_2}^{e_1}\frac{udu}{y}=3.141592903870775$, 
$\int_{e_3}^{e_2}\frac{udu}{y}=20.4095744i$ etc. In this case we obtain
$2\pi (7.96669\times 10^{-8}){\rm radians/revolution}=42.8689 {\rm \frac{arc-sec}{century}}$ in conflict with observations.
For $\Lambda=10^{-56}{\rm cm^{-2}}, {\cal L}= 1.1848820116975453\times 10^{-28}\;{\rm
cm^{-2}}\;{s^{2}}\;{\rm cm^4}\;{s^{-2}},
E=0.0299792454178\times 10^{12}{\rm cm\; s^{-1}}$ the roots are:
\begin{eqnarray}
e_1&=&0.0000000105747129978\times \frac{4}{\alpha_s} \nonumber \\
e_2&=&0.0000000160482602833\times \frac{4}{\alpha_s} \nonumber \\
e_3&=&0.2499999733770\times \frac{4}{\alpha_s} \nonumber \\
e_4&=&(-5.5897098056\times
10^{-32}-2.669498120170\times10^{-20}i)\times \frac{4}{\alpha_s}
\nonumber \\
e_5&=&\bar{e}_4
\end{eqnarray}
with  the real half period: $\int_{e_2}^{e_1}\frac{udu}{y}=3.141592904505435$, which
leads to $\Delta_{\omega}^{GTR}=42.9776 {\rm \frac{arc-sec}{century}}$ in agreement 
with observations (\ref{Mercobs}).

\begin{table}
\begin{center}
\begin{tabular}{|c|c|c||c|}\hline\hline
. & $\Lambda=0$ & $\Lambda=-10^{-42}{\rm cm^{-2}}$
&$\Lambda=10^{-42}{\rm cm^{-2}}$ \\
\hline\hline
$\omega$ & 3.141592904505435 & 3.14159290255 & 3.141592903870775  \\
\hline
$\Delta_{\omega}^{GTR}$ & $42.9776{\rm \frac{arcsec}{century}}$
&$42.6427{\rm \frac{arcsec}{century}}$ & $42.8689{\rm
\frac{arcsec}{century}}$ \\
\hline
$r_{P}$ & $4.60057668\times 10^{12}{\rm cm}$ & $4.60057668\times 10^{12}{\rm
cm}$ & $4.60057668\times 10^{12}{\rm cm}$ \\
\hline
$r_{A}$ & $6.98186816 \times 10^{12}{\rm cm}$ & $6.98186817\times 10^{12}{\rm
cm}$ & $6.98186815 \times 10^{12}{\rm cm}$ \\
\hline \hline
\end{tabular}
\end{center}
\caption{Theoretical predictions for perihelion precession for
$\Lambda\not=0$, and 
${\cal L}= 1.1848820116975453\times 10^{-28}\;{\rm
cm^{-2}\;{s^{2}}},
E=0.0299792454178\times 10^{12}{\rm cm\; s^{-1}}$. For comparison we
list the $\Lambda=0$ case for the same set of values for ${\cal L},E$.
For this choice of ${\cal L}$, $L^2_M=7.36080550\times 10^{38}
{\rm cm^4 s^{-2}}$.}
\label{EINSTEINGVKSBW1}
\end{table}

\begin{table}
\begin{center}
\begin{tabular}{|c|c|c||c|}\hline\hline
. & $\Lambda=-10^{-55}{\rm cm^{-2}}$ & $\Lambda=10^{-56}{\rm cm^{-2}}$
&$\Lambda=0$ \\
\hline\hline
$\omega$ & 3.14159290450544 & 3.141592904505435 & 3.141592904505435 \\
\hline
$\Delta_{\omega}^{GTR}$ & $42.9776{\rm \frac{arcsec}{century}}$
&$42.9776{\rm \frac{arcsec}{century}}$ & $42.9776{\rm
\frac{arcsec}{century}}$ \\
\hline
$r_{P}$ & $4.60057668\times 10^{12}{\rm cm}$ & $4.60057668\times 10^{12}{\rm
cm}$ & $4.60057668\times 10^{12}{\rm cm}$ \\
\hline
$r_{A}$ & $6.98186816 \times 10^{12}{\rm cm}$ & $6.98186816\times 10^{12}{\rm
cm}$ & $6.98186816 \times 10^{12}{\rm cm}$ 
 \\
\hline \hline
\end{tabular}
\end{center}
\caption{Theoretical predictions for perihelion precession for
$\Lambda\not=0$, and 
${\cal L}= 1.1848820116975453\times 10^{-28}\;{\rm
cm^{-2}\;{s^{2}}},
E=0.0299792454178\times 10^{12}{\rm cm\; s^{-1}}$. For comparison we list 
the corresponding results for the $\Lambda=0$ case.
For this choice of ${\cal L}$, $L^2_M=7.36080550\times 10^{38}{\rm cm^4
s^{-2}}$.}
\label{EINSTEINGVKSBW2}
\end{table}

We summarise our results in Table \ref{EINSTEINGVKSBW1} and \ref{EINSTEINGVKSBW2}.
We also repeated our calculations for the values for ${\cal L},E$ as in
Tables 1 and 2, and we found: For the 
values for ${\cal L},E$ as in table 1 for $\Lambda=-10^{-55}{\rm cm
^{-2}}$, $\int_{e_2}^{e_1}\frac{udu}{y}=3.1415929045292983$,
$\Delta_{\omega}^{GTR}=42.9817 {\rm{\frac{arc-sec}{century}}}$ and for 
$\Lambda=10^{-56}{\rm cm ^{-2}}$, $\int_{e_2}^{e_1}\frac{udu}{y}
=3.141592904529299$ and
$\Delta_{\omega}^{GTR}=42.9817{\rm{\frac{arc-sec}{century}}}$. 

For the values for ${\cal L},E$  chosen  in table 2, 
for $\Lambda=-10^{-55}{\rm cm ^{-2}}$, 
$\int_{e_2}^{e_1}\frac{udu}{y}=3.14159290452253$,
$\Delta_{\omega}^{GTR}=42.9805{\rm{\frac{arc-sec}{century}}}$, while
for $\Lambda=10^{-56}{\rm cm ^{-2}}$, $\int_{e_2}^{e_1}\frac{udu}{y}
=3.141592904524534$ and  $\Delta_{\omega}^{GTR}=42.9809{\rm{\frac{arc-sec}{century}}}$.
Thus we see that the cosmological constant depending on its magnitude 
has an effect on the
perihelion precession and in fact is in conflict with observations for
magnitudes $|\Lambda| \sim 10^{-42}{\rm
cm^{-2}}$ and larger
while current favoured values of the cosmological constant
\cite{PERLMUTTER,BAHCALL,DEBERNAR} are
compatible with observations of the Mercury's perihelion advance and
do not have significant effect either on the perihelion precession
or the radii $r_P,r_A$.
Evidently, more precise observations can put a rigorous upper bound on
the magnitude 
of the cosmological constant.

\section{Discussion and further applications}

In this paper we dealt with the motion of a test particle in a central 
gravitational field.
In particular we investigated the orbit of  planet Mercury around the 
gravitational Schwarzschild field of the Sun by solving exactly the 
geodesic differential equations 
with and without the cosmological constant and determined  the perihelion 
advance. 
As we saw in the main body of the paper, the analytic solution of the orbital 
problem is provided by the solution of the inversion problem first enunciated 
by Abel and Jacobi in their study of elliptic and hyperelliptic modular 
functions.

The perihelion precesion of the planet in the exact solutions 
obtained depends on fundamental properties of the modular functions and 
in particular their periods.

We have also calculated the precise perihelion and aphelion distances of
the planet. The calculated theoretical values 
compare favourably 
with best current values for the orbital data for Mercury \cite{NASA}.

For zero cosmological constant the exact solution for the orbit of Mercury 
is expressed by the Weierstra$\ss$ Jacobi modular form.
The orbital data of the planet are reproduced for positive
discriminant ($\Delta>0$) and when all roots of the cubic are real.
For the particular choice of the free parameters given in
Tables \ref{EINSTEIN1}-\ref{EINSTEIN3} the theoretical predictions for 
the perihelion 
advance are in agreement with observations Eq.(\ref{Mercobs}). 
The complex structure of the torus that describe the particular orbital 
solutions is $\tau=6.497\;i$. We note that the modular 
properties of the exact solutions in large-scale cosmology including 
cosmological constant effects allow for both signs of the discriminant
$\Delta$ and in some interesting cases the complex structure of the corresponding tori is a fixed point of the modular group \cite{GVKSBW}.

For non-zero cosmological constant the exact solution for the 
orbit is provided by quotients of genus two theta functions and
therefore is described by a hyperelliptic curve of genus two. 
Alternatively, the solution can be reexpressed in terms of generalized
Weierstra$\ss$ functions.
We investigated the effect of various values for the cosmological 
constant on the perihelion precession. Magnitudes of $\Lambda \sim 
10^{-42}cm^{-2}$  are in conflict with observations (\ref{Mercobs}),
while current favoured  values from large scale cosmology are
compatible with observations. More accurate measurements will place a 
rigorous upper bound on the magnitude of the cosmological constant.

The results of this paper, are in full agreement with the method
suggested in \cite{GVKSBW} namely: the properties of modular 
theta functions associated with Riemann surfaces and in particular
the non-linear differential equations they obey, provide a new way to 
deal with Einstein' equations. In this way, the non-linearity of the 
gravitational field equations can be tackled effectively and general 
exact solutions be generated.

Our exact treatment of the geodesic equations can be applied to a 
variety of problems 
in cosmology where the two body central orbit problem in general 
relativity is encountered. More specifically our  
techniques can be applied to address  problems such as : 
the determination of
orbital parameters of stars around massive gravitational objects such as
Neutron Stars \cite{Damour} or candidate Black Holes \cite{Schodel}, galactic 
dynamics including
the missing mass problem \cite{GVKSBW1} which will be a subject of 
separate publication , or other scales in cosmology \cite{Roberts, MAEFLP}.
It would also be interesting to generalise the techniques developed in
this paper to the 
gravitational field of a rotating mass.  Thus effects, like the
Lense-Thirring \cite{LTPr} precession associated with such
gravitational fields
could be investigated.
We note that the Schwarzschild solution because of its static
character does not describe rotation of the mass distribution. 
Such generalisation is of additional interest for the following
reason: taking into account very fine corrections due to a non-zero
cosmological constant calculated in this paper,  one should 
look carefully if they are not
comparable with other subtle relativistic effects, like the
rotation of the Sun and the associated Lense-Thirring effect. Other subtle effects include the 
constant loss of mass during cosmological times.

In addition, 
the precise general relativistic predictions 
obtained can be compared
with the predictions of scalar-tensor 
type gravity theories such as the Brans-Dicke theory. We already 
mentioned that the latter theory for small values of its 
additional parameter $\omega$ ($\omega \sim 5$) is in conflict with 
the observations for the perihelion precession of Mercury Eq.(\ref{Mercobs})
and current estimations of the Sun's quadrupole moment \cite{Pireaux,Will}.

Non-commutative field theories are also strongly constrained from the
precession of the perihelion of  Mercury \cite{Short}.

We are entering a new era in cosmology where precise theoretical results 
can be compared with the ever increasing accuracy of observations.

\section*{Acknowledgements} We are gratefull to D. L$\rm{\ddot
u}$st for proof-reading the paper and J. Sanchez for useful
correspondence. 
The work of GVK was supported by a 
String theory, Cosmology Project Nr.3140 1818 Leibniz-Preis, and a
Graduiertenkolleg research fellowship.
We are gratefull to the Referees for their  constructive remarks and 
criticisms that helped to improve the presentation.

\appendix

\section{Definitions of genus-2 theta functions that solve Jacobi's inversion 
problem}

Riemann's theta function \cite{BerGeorgR} for genus $g$ is defined as follows:

\begin{equation}
\Theta(u):=\sum_{n_1,\cdots,n_g} e^{2\pi i u n+i \pi \Omega n^2}
\end{equation}
where $\Omega n^2:=\Omega_{11} n_1^2+\cdots 2 \Omega_{12}n_1 n_2 +\cdots$ 
 and $un:=u_1n_1+\cdots u_g n_g$. The symmetric $g\times g$ complex matrix 
$\Omega$ whose imaginary part is positive definite is a member of the 
set called Siegel upper-half-space denoted as ${\cal L}_{S_g}$. It is 
clearly the generalization of the ratio of half-periods $\tau$ in the genus $g=1$ case.    
For genus $g=2$ the Riemann theta function can be written in 
matrix form:

\begin{eqnarray}
\Theta(u,\Omega)&=&\sum_{{\bf{n}} \in Z^2} e^{\pi i {\bf ^{t} n}\Omega {\bf n}+
2\pi i {\bf ^{t} n} {\bf u}} \nonumber \\
&=& \sum_{n_1,n_2} e^{\pi i \left(\begin{array}{cc}n_1 & n_2\end{array}\right) \left(\begin{array}{cc}
\Omega_{11} & \Omega_{12} \\
\Omega_{12} &  \Omega_{22} \end{array}\right) \left(\begin{array}{c}
n_1 \\
n_2\end{array}\right) +2 \pi i \left(\begin{array}{cc} n_1 &n_2 \end{array}\right) \left(\begin{array}{c}
u_1 \\
u_2\end{array}\right)} \nonumber \\
\end{eqnarray}
Riemann's theta function with characteristics is defined by:
\begin{equation}
\Theta(u;q,q^{\prime}):=\sum_{n_1,\cdots,n_g}e^{2\pi i u(n+q^{\prime})+
i\pi \Omega (n+q^{\prime})^2+2 \pi i q(n+q^{\prime})}
\label{thechara}
\end{equation}
herein $q$ denotes the set of $g$ quantities $q_1,\cdots,q_g$ and $q^{\prime}$ 
denotes the set of $g$ quantities $q^{\prime}_{1},\cdots,q^{\prime}_g$.
Eq.(\ref{thechara}) can be rewritten in a suggestive matrix form:
\begin{equation}
\Theta \left [\begin{array}{c} q^{\prime} \\ q \end{array} \right](u,\Omega)=
\sum_{n \in Z^g} e^{\pi i ^{t} (n+q^{\prime}) \Omega (n+q^{\prime})+
2 \pi i ^{t} (n+q^{\prime})(u+q)}, \;\;\;\;q,q^{\prime} \in Q^g
\end{equation}

The theta functions whose quotients provide a solution to Abel-Jacobi's inversion 
problem are defined as follows \cite{BAKER}:
\begin{equation}
\theta(u;q,q^{\prime}):=\sum e^{au^2+2 hu(n+q^{\prime})+b(n+q^{\prime})^2+
2 i\pi q(n+q^{\prime})}
\end{equation}
where the summation extends to all positive and negative integer values of 
the $g$ integers $n_1,\cdots,n_g$, $a$ is any symmetrical matrix whatever 
of $g$ rows and columns, $h$ is any matrix whatever of $g$ rows and columns, 
in general not symmetrical, $b$ is any symmetrical matrix whatever of $g$ 
rows and columns, such that the real part of the quadratic form  $bm^2$ 
is necessarily negative for all real values of the quantities $m_1,\cdots,
m_g$, other than zero, and $q,q^{\prime}$ constitute the characteristics of 
the function. The matrix $b$ depends on $\frac{1}{2}g (g+1)$ independent 
constants; if we put $i\pi \Omega=b$ and denote the $g$-quantities 
$hu$ by $i\pi U$, we obtain the relation with Riemann's theta function:
\begin{equation}
\theta(u;q,q^{\prime})=e^{au^2}\Theta(U;q,q^{\prime})
\end{equation}

The dependence of genus-2 theta functions on two complex variables is
denoted by:  $\theta(u;q,q^{\prime})=\theta(u_1,u_2;q,q^{\prime})$,
the dependence on the Siegel moduli matrix $\Omega$ by:
$\theta(u_1,u_2,\Omega;q,q^{\prime})$.
To every half-period one can associate a set of characteristics.
For instance, the period $u^{a,a_1}=\frac{1}{2}\left(\begin{array}{cc}
1 & 0 \\
1 & 0 \end{array}\right)$ while $\theta(u)$ is a theta function of 
two variables with  zero
characteristics, i.e. $\theta(u)=\theta(u;0,0)=\theta \left
[\begin{array}{cc} 0 & 0 \\ 0 & 0 \end{array} \right](u,\Omega)$.
Also,
Weierstra$\ss$ had associated a symbol for each of the six odd theta
functions with characteristics 
and the ten even theta functions of genus two. For example,
$\theta(u)$ is associated with the Weierstra$\ss$ symbol 5 or
occasionaly the number appears as a subscript, i.e. $\theta(u)_5$.

The matrix elements $h_{ij},\Omega_{ij}$ can be explicitly written in
terms of the half-periods $U_r^{x,a}$ defined in section 3.6.
For instance, the matrix element $h_{11}=\frac{U_2^{e_4,e_3}}{2(
U_1^{e_4,e_3}U_2^{e_2,e_1}-U_1^{e_2,e_1}U_2^{e_4,e_3})}\times \pi i$, while 
$\Omega_{11}=\frac{U_1^{e_1,e_0}U_2^{e_4,e_3}-U_2^{e_1,e_0}U_1^{e_4,e_3}}{
U_2^{e_2,e_1}U_1^{e_4,e_3}-U_1^{e_2,e_1}U_2^{e_4,e_3}}$.

Let us perform some consistency checks for particular cases of the inversion problem:
\begin{eqnarray}
\int_{a_1}^{u_1} \frac{du}{y}+\int_{a_2}^{u_2} \frac{du}{y}\equiv \phi_1 \nonumber \\
\int_{a_1}^{u_1} \frac{u\;du}{y}+\int_{a_2}^{u_2} \frac{u\;du}{y} \equiv \phi_2
\label{Umkehr1}
\end{eqnarray}
For instance for
$\phi_1=\phi_1^{c_2,a_1}, \phi_2=\phi_2^{c_2,a_1}$, 
$u_2=a_2=e_1, u_1=c_2=e_2$. Let us check if the formulas
eqs.(\ref{inveb})-(\ref{inve2}) reproduce the chosen upper limits of
the period integrals in the inversion problem.

In this case, 
\begin{eqnarray}
\frac{1}{A_1}\frac{\theta^2\left(\phi|\phi^{a_1,a}\right)}{\theta^2(\phi)}
&=&(u_1-a_1)(u_2-a_1)=(c_2-a_1)(a_2-a_1) \nonumber \\
\frac{1}{A_2}\frac{\theta^2\left(\phi|\phi^{a_2,a}\right)}{\theta^2(\phi)}
&=&0
\end{eqnarray}
and therefore using (\ref{inve2}) we obtain $u_2=a_2$. Then from 
\begin{equation}
u_1=a_1 +\frac{1}{A_1
(u_2-a_1)}\frac{\theta^2(\phi|\phi^{a_1,a})}{\theta^2(\phi)}=a_1+\frac{1}{a_2-a_1}(c_2-a_1)(a_2-a_1)=c_2.
\end{equation}

\section{Roots of the cubic and special orbit cases for vanishing cosmological constant}

The roots of the elliptic curve can be calculated analytically using the 
algorithm developed by Tartaglia and Cardano \cite{TARTACARDA}.
Their general expressions are given by:
\begin{eqnarray}
r_1&=& \frac{2^{1/3}\left\{2+18{\cal L}c^2-27{\cal L}E^2+3\sqrt{3}
\sqrt{\delta}\right\}^{2/3}+2-6{\cal L}c^2}
{12\; 2^{2/3}\left\{(2+18 {\cal L}c^2-27{\cal L} E^2)+3\sqrt{3}
\sqrt{\delta}\right\}^{1/3}} \nonumber \\
r_2&=&\frac{\rho 2^{1/3}\left\{2+18{\cal L}c^2-27{\cal L}E^2+3\sqrt{3}
\sqrt{\delta}\right\}^{2/3}+\rho^2 (2-6{\cal L}c^2)}
{12\; 2^{2/3}\left\{(2+18 {\cal L}c^2-27{\cal L} E^2)+3\sqrt{3}
\sqrt{\delta}\right\}^{1/3}} \nonumber \\
r_3&=&\frac{\rho^2 2^{1/3}\left\{2+18{\cal L}c^2-27{\cal L}E^2+3\sqrt{3}
\sqrt{\delta}\right\}^{2/3}+\rho (2-6{\cal L}c^2)}
{12\; 2^{2/3}\left\{(2+18 {\cal L}c^2-27{\cal L} E^2)+3\sqrt{3}
\sqrt{\delta}\right\}^{1/3}}
\end{eqnarray}
where
\begin{equation}
\delta:={\cal L}\left(8c^4{\cal L}+4{\cal L}^2c^6+c^2(4-36E^2{\cal L})+E^2
(27E^2{\cal L}-4)\right)
\end{equation}
and $\rho=e^{2\pi i/3}$.

Equation (\ref{orbitnol}), represents the exact solution in closed 
analytic form for the 
orbit of a test particle in the central gravitational field of the
Sun assuming zero cosmological constant. As it was mentioned in
section 3.1, not  all  choices of initial conditions will lead to
planetary orbits. 

Below we list the orbits that
correspond to specific choices of initial conditions and in particular 
to the two elliptic curves for which one of their Weierstra$\ss$ invariants 
vanishes.

i) Case of negative discriminant and $g_2=0,g_3\not =0$.

In this case for ${\cal L}=\frac{1}{3c^2}$ or equivalently,
$L^2_M=3c^2 \alpha_S^2$, we obtain:
$g_2=0,\;g_3=\frac{1}{216}-\frac{1}{144}-\frac{1}{48c^2}(E^2-c^2)$.
The discriminant $\Delta=-27g_3^2<0$ and two roots are complex
conjugates and one is real. The Weierstra$\ss$ function $\wp(\phi)$ is real along the diagonal of the fundamental period parallelogram (FPP), and the real 
root $e_2$ is located at the point $\omega+\omega^{\prime}$. As the 
argument of the Weierstra$\ss$ function changes from $\omega+\omega^{\prime}$ to $0$ 
along the diagonal of FPP, its value changes from $e_2$ to $\infty$.

For $E=\sqrt{\frac{8}{9}}c$, also $g_3=0$ and then there is no elliptic curve 
since the discriminant vanishes and all three roots are zero.

ii) Case of $g_3=0,g_2 >0$.

In this case there are three real roots, one of them $e_2=0$ and the other two 
are equal in magnitude and opposite in sign, 
$e_1=\frac{\sqrt{1-3 c^2 {\cal L}}}{4 \sqrt{3}},e_3=-e_1$. Here, as the argument of 
Weierstra$\ss$ function $\wp$, travels along the line from $\omega+\omega^{\prime}$ to 
$\omega$ in the FPP, its value changes from $0$ to $e_1>0$. Thus, the radius 
vector varies from $r_{e_2}=3 \alpha_S$ to $r_{e_1}=\frac{\alpha_S}{4
e_1+\frac{1}{3}}$.  On the other hand in the region, of the FPP along
the line from $\omega+\omega^{\prime}$ to $\omega^{\prime}+2\omega$
the Weierstra$\ss$ function is negative $e_3\leq \wp(\phi)\leq 0$. The
radius vector increases  from $3 \alpha_S$ and tends asymptotically to
infinity when $4\wp(\phi)+\frac{1}{3}\rightarrow 0$. 

iii) Case of vanishing discriminant.

The discriminant $\Delta=g_2^3-27 g_3^2$ can be written as follows:
\begin{equation}
\Delta=\frac{{\cal L} \left(-4 {\cal L}^2 c^6+{\cal L}(-
27E^4+36 c^2 E^2-8c^4)+4(E^2-c^2)\right)}{256}
\end{equation}

It vanishes, besides the case we
mentioned above for which $g_2=g_3=0$, also for ${\cal L}=0$ and for 
$${\cal L}=-\frac{1}{8c^6}\left(8c^4-36c^2 E^2+27 E^4 
\pm \sqrt{-512c^6E^2+1728c^4 E^4-1944c^2E^6+729E^8}\right)$$

\end{document}